\documentclass[journal=jpca,manuscript=perspective,layout=twocolumn]{achemso}

\usepackage[fontsize=10pt]{fontsize}
\usepackage{chemformula} % Formula subscripts using \ch{}
\usepackage[T1]{fontenc} % Use modern font encodings
\usepackage{crimson}
\usepackage{hyperref}
\usepackage{xurl}

\hypersetup{
    colorlinks=true,
    linkcolor=blue,
    citecolor=blue,
    filecolor=blue,
    urlcolor=blue,
}

\author{Jonathan E. Moussa}
\email{godotalgorithm@gmail.com}
\affiliation{Molecular Sciences Software Institute, Virginia Tech, Blacksburg, Virginia 24060, United States}

\title{The Enduring Relevance of Semiempirical Quantum Mechanics} %: A Requiem

\abbreviations{IR,NMR,UV}
\keywords{American Chemical Society, \LaTeX}

\begin{document}

\begin{tocentry}

\includegraphics{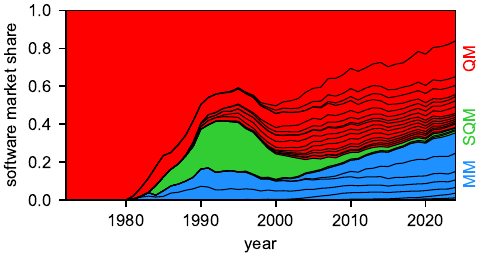}

\end{tocentry}

\begin{abstract}
The development of semiempirical models to simplify quantum mechanical descriptions of atomistic systems is a practice that started soon after the discovery of quantum mechanics
 and continues to the present day.
There are now many methods for atomistic simulation with many software implementations and many users, on a scale large enough to be considered as a software market.
Semiempirical models occupied a large share of this market in its early days, but the research activity in atomistic simulation has steadily polarized over the last three decades towards
 general-purpose but expensive \textit{ab initio} quantum mechanics methods and fast but special-purpose molecular mechanics methods.
I offer perspective on recent trends in atomistic simulation from the middle ground of semiempirical modeling, to learn from its past success and consider its possible paths to future growth.
In particular, there is a lot of ongoing research activity in combining semiempirical quantum mechanics with machine learning models and some unrealized possibilities of tighter integration
 between \textit{ab initio} and semiempirical quantum mechanics with more flexible theoretical frameworks and more modular software components.
\end{abstract}

%S1
\section{Introduction}

%P1.1 - Early history of SQM
Accurate \textit{ab initio} quantum mechanics (QM) calculations are expensive, and scientists' desire to perform such calculations predates the availability of computers capable of performing them.
To resolve this discrepancy, scientists built simple approximate models of these calculations by hand and on early computers.
These scientists often tuned parameters to match experimental data and increase the physical relevance of their models, and from this activity three semiempirical quantum mechanics (SQM) traditions emerged.
The most prolific of these traditions occurred in quantum chemistry, starting from H\"{u}ckel's model of $\pi$ orbitals in planar hydrocarbons \cite{huckel}
 and Hoffmann's extension to $\sigma$ orbitals for a more general description of hydrocarbons \cite{hoffmann}.
They inspired the development of more general thermochemistry models of organic molecules such as Pople's CNDO \cite{CNDO} and INDO \cite{INDO} models
and Dewar's MINDO/3 \cite{MINDO3} and MNDO \cite{MNDO} models, before reaching a pinnacle (by citation count) in the AM1 model \cite{AM1}.
H\"{u}ckel's work in chemistry was directly inspired by Bloch's use of atomic orbitals (AOs) to describe electronic energy bands in crystals \cite{bloch},
 which also inspired the AO-based Slater--Koster tight-binding formalism in physics \cite{slater_koster}.
While most SQM models have been constructed in an AO basis, the empirical pseudopotential method \cite{epm} (EPM)
 uses pseudopotentials \cite{pseudopotentials} to enable the efficient approximation of atomic orbitals in a plane-wave basis,
 which has a popular parameterization for binary semiconductors \cite{chelikowsky_cohen}.
Despite their common motivations, the development and application of these three SQM traditions has been largely independent, as shown in Table \ref{sqm_crosscite}.

%Table 1 - SQM citation correlations
\begin{table}[b]
  \caption{Number of citations and joint citations to the most cited papers of the three SQM traditions, as estimated using Google Scholar.}
  \label{sqm_crosscite}
  \begin{tabular}{l | lll}
    \hline
    & AM1 & Slater--Koster & EPM \\
    \hline
    AM1\cite{AM1} & 16,800 & 36 & 0 \\
    Slater--Koster\cite{slater_koster} & & 6,690 & 99 \\
    EPM\cite{chelikowsky_cohen} & & & 2,790 \\
    \hline
  \end{tabular}
\end{table}

%Figure 1 - Software citations
\begin{figure*}[!t]
\includegraphics{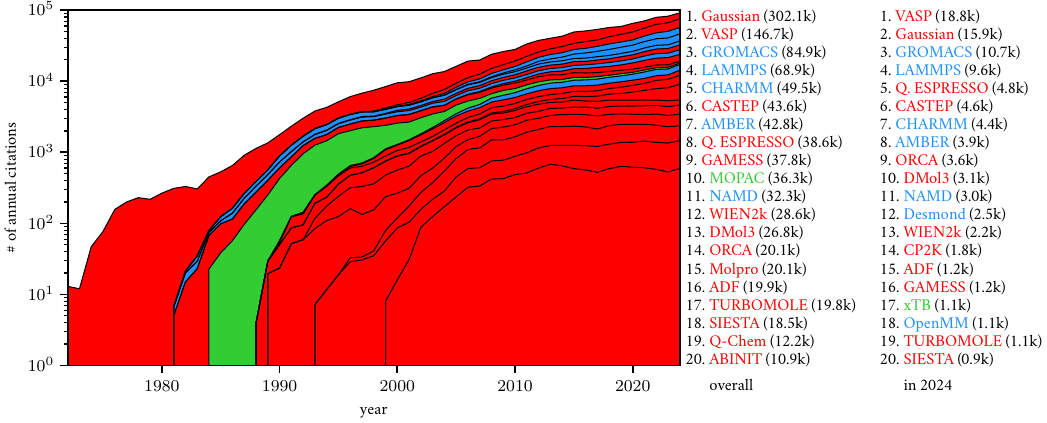}
     \caption{Number of citations to the 20 most cited atomistic simulation engines, as estimated using Google Scholar. Color denotes the software's primary simulation type: QM (red), SQM (green), or MM (blue). The history of aggregate annual citations is plotted on a logarithmic scale with its components ordered by overall citation count.
}\label{software_citation}
\end{figure*}

%P1.2 - Modern trends in atomistic simulation methods & software
While SQM method development has continued to the present day, it has been overshadowed in accuracy by \textit{ab initio} QM methods and in efficiency by molecular mechanics (MM) methods.
The growing abundance of computational resources has enabled the widespread use of QM methods,
 and the efficiency needs of large-scale atomistic simulations have only been satisfied by MM methods.
In general, a polarization of method development in atomistic simulation between accurate QM methods and fast MM methods has been driven by the pursuit of specific challenging applications
 that demand either accuracy or efficiency beyond what computational methods have been able to deliver.
In quantum chemistry, this polarization \cite{coulson,pople_perspective,karplus_perspective} was driven by the pursuit of a fully predictive QM-based understanding of small molecules
 and a more descriptive atomistic understanding of biological processes by embedding small QM regions in large MM simulations and coupling them with QM/MM methods \cite{qmmm}.
Modern QM method development has focused on density functional theory \cite{dft} (DFT) to provide a baseline level of accuracy that might then
 be refined by more expensive quantum many-body methods such as coupled-cluster theory \cite{cc} or many-body Green's functions \cite{green}.
Although the AM1 model paper is among the most cited scientific papers, several important DFT methodology papers collectively have an order of magnitude more citations \cite{AM1_cite}.
Practical DFT calculations retain several important SQM influences: the use of semiempirical density functionals such as B3LYP \cite{b3lyp},
 the evolution of EPM models into the plane-wave pseudopotential formalism widely used in physics \cite{modern_pseudopotentials},
 and the generation of system-specific tight-binding models as a common post-processing step of plane-wave pseudopotential calculations \cite{wannier90}.
Despite these influences, a large gap in computational cost continues to separate QM from SQM methods, and a similarly large gap separates SQM from MM methods.
For the elementary task of atomic force evaluation, the relative computational cost separating QM, SQM, and MM calculations is typically larger than one thousand \cite{linear_scaling2}.
Thus, I use the QM/SQM/MM categorization in this paper to denote regimes of computational cost necessitated by the level of theory.

%P1.3 - Supply vs. demand perspective
Both the number of scientists using atomistic simulation and the amount of computational resources available per scientist have been increasing exponentially with time over the last five decades,
 and simulation methods now have the largest impact when implemented in software with a large, active user base.
The development and use of this software now occurs on a large enough scale that it can be considered in economic terms as a market with price and quantity governed by supply and demand.
The supply of atomistic simulation software is generated by both academic and commercial software developers in a quantity that increases with price,
 where ``price'' is the amount of grant money that can be acquired for academic software development or the amount of money that a company can charge for a commercial software license.
One relevant ``quantity'' is the amount of publicly available software, and I estimate the number of atomistic simulation engines to be 167 for QM, 33 for SQM, and 120 for MM.
On the demand side, the ``price'' of this software is the price of a software license plus the computational cost to run it and the human time and expertise to operate it.
For 74\% of these atomistic simulation engines, the license price is zero because of the trend in academic software development towards free and open-source software (FOSS) licenses \cite{foss,foss_chemistry}.
Following a recent study \cite{software_trends}, I estimate software citations to popular atomistic simulation engines in Fig.\@ \ref{software_citation} as a ``quantity'' of academic demand that
 has grown in time with the increasing abundance of human and computer resources.
An important but even more difficult to quantify component of this market is industrial demand, including the commercial software companies that profit from this activity
 and their ability to reinvest profits back into method and software development to advance the field with resources beyond public funding for basic research.
As recently suggested \cite{market_analysis}, a practical measure of this commercial market is the size of companies that it can support,
 and a natural point of reference is the larger but related market for continuum simulation software.
The continuum simulation software market supports multiple large, public companies such as Dassault Syst\`{e}mes, Synopsys, Autodesk, and Cadence Design Systems
 that each currently employ over 10,000 people.
The only public company currently operating entirely within the atomistic simulation market is Schr\"{o}dinger,
 which sells software with QM, SQM, and MM features and supports approximately 900 employees.
Companies focusing on QM software tend to be much smaller, with Gaussian and VASP developed by eponymous companies that each support approximately 20 employees.
Company size is a very different metric from software citations, as evidenced by Gaussian and VASP being more highly cited than any software sold by Dassault, Synopsys, Autodesk, or Cadence,
 and both may have some correlation with industrial demand.

%P1.4 - More thoughtful rationale for SQM
SQM software had an academic market share greater than 20\% in the 1990s that has since collapsed to around 1\%.
Relative to QM and MM methods and software, the supply of and demand for SQM methods and software are now very low.
However, the past successes of SQM may contain important lessons for QM now, and the ongoing successes of QM may be important lessons for SQM in the pursuit of a more successful future.
Here I present several perspectives on how SQM and QM may benefit from each other, which is intended to complement an SQM roadmap article that I recently contributed to \cite{sqm_roadmap}.
The first perspective is a deconstruction of SQM models into elementary components and a consideration of how new SQM models might be constructed
 from the many components available in modern QM methods and software.
The second perspective is a proposal for how SQM and QM could be more unified by Hamiltonian model forms that encode the simple structures that have enabled SQM's low computational cost
 while being general enough for \textit{ab initio} QM in an arbitrarily large basis.
The third perspective is an overview of some recent trends in SQM model development and some suggestions drawing from the technical successes of QM.
I believe that the atomistic simulation market would greatly benefit from a broader range of simulation capabilities of varying cost, accuracy, and breadth of capability
 from more unified SQM and QM development that has a stronger emphasis on the mid-range cost segment of the market between MM and QM.

%S2
\section{Models from components}

\begin{figure}[!t]
\includegraphics{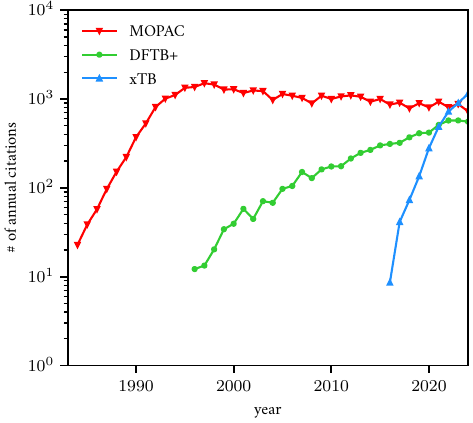}
     \caption{Number of citations to the 3 most cited SQM software programs, as estimated using Google Scholar.
     Note that the DFTB+ estimate effectively includes an estimate of the DFTB software that preceded it prior to 2007.
}\label{sqm_citation}
\end{figure}

%P2.1 - Summary of most popular SQM models & software
Historically, the most popular SQM software has been MOPAC \cite{mopac}, which implemented the MNDO family \cite{MNDO} of SQM models including AM1 \cite{AM1}.
The popularity of DFT in the 1990's inspired the density functional tight binding (DFTB) method \cite{dftb} and eventually the DFTB+ software \cite{dftbplus}.
More recently, the Grimme group at the University of Bonn developed the GFN family of SQM models and its software implementation in xTB \cite{xtb}.
These programs have been the primary platforms for the development and use of their respective SQM models, and I visualize their citation data in Fig.\@ \ref{sqm_citation}.
Other impactful examples of SQM models and software are Zerner's INDO model and ZINDO program \cite{zindo},
 the Naval Research Laboratory (NRL) tight-binding model \cite{nrltb}, and the NEMO tight-binding model and program \cite{nemo}.

%Figure 2 - SQM software citations

%P2.2 - Computation-limited tasks vs. exhaustion of computational resources
While not a strict separation, SQM software is typically intended for specific computational tasks with a limited cost while QM software
 is typically intended to exhaust available computing resources to achieve as much certainty as possible, as exemplified by Pople's diagrams \cite{pople_diagrams} and Perdew's ladder \cite{perdew_ladder}.
For example, MOPAC predicts the heat of formation and equilibrium geometry of molecules, xTB is heavily used with the CREST software \cite{crest} for conformational sampling of molecules,
 ZINDO predicts the electronic spectra of molecules, and NEMO is used for semiconductor device modeling,
 while DFTB+ and the NRL tight-binding model are more generically used as low-cost approximations of DFT calculations.
Both QM and SQM software have numerical parameters that can be tuned, but only QM software effectively enables a systematic reduction of all sources of error if sufficient computational resources are available.
While the computational cost of individual MM calculations is much lower than QM and SQM calculations, MM software is typically used for tasks that require a very large number of calculations
 to control bias and reduce uncertainty of statistical estimates, and available computing resources are often exhausted to that end.
Similarly, SQM software provides the most cost-effective access to large quantities of electronic properties that are not provided by MM calculations,
 for ensemble averaging as in CREST or for high-throughput virtual screening of molecules and materials \cite{screening}.

%P2.3 - QM software as a large collection of features & options
Mature QM software tends to have a large number of features and options and a broad domain of applicability.
However, the practical use of QM software tends to cluster into specific choices of features and options (e.g.,\@ basis set and density functional) based on popularity and precedence in the literature.
Often the rationale and justification of these choices traces back to the original papers from the method or software developers that first described the implementation and appropriate use of a feature.
With their great flexibility, QM software can be used in a more low-cost mode of operation,
 but that often comes at a significant cost of human time in calibrating options and building confidence that the desired degree of reliability and accuracy can be achieved.
In addition to their low computational cost, SQM software can also save some of the human time spent on calibration because SQM models have been optimized for specific tasks
 and articulate their accuracy by a closeness of fit to reference data sets that are representative of their intended use.

%P2.4 - Assembling QM components into new SQM models
To the degree that QM methods and software are modular, they can be considered as components for the construction of new SQM models.
In this way, some low-cost use of QM software could avoid the human cost of calibration for common atomistic simulation tasks.
Indeed, this has been a popular activity in modern QM method development with particularly notable results from Grimme
 such as semiempirical density functionals \cite{semiempirical_dft}, the D$x$ models \cite{d_models} for dispersion corrections to popular density functionals,
 and the $x$-3c models \cite{3c_models} that correct both basis-set incompleteness and correlation errors.
However, in using components from modern QM software, these SQM models do not have access to the low-cost components that enable the efficiency of dedicated SQM software.
The efficiency of AO-based SQM software depends on the neglect of diatomic differential overlap (NDDO) approximation,
 which is an approximate factorization of four-center Coulomb integrals that was introduced during the development of the CNDO model \cite{nddo} but still has no modern QM counterpart.
Similarly, the efficiency of EPM models depends on very soft pseudopotentials that enable a very small plane-wave basis with a very low energy cutoff,
 which was lost in the transition towards harder, more transferable pseudopotential forms in modern plane-wave pseudopotential QM methods.

%P2.5 - Tightly coupled stack
The ability for QM software to be reshaped into new SQM models depends on how effectively it has been decomposed into reliable and reconfigurable components,
 which is a steadily growing emphasis of atomistic simulation software development \cite{components}.
Such components are a tightly coupled stack of interfaces (e.g.,\@ user interface, application programming interface, input and output data formats) on top of software, algorithms, and methods,
 resting on foundations of theoretical formalism.
A substantial amount of recent activity has focused on the interface and software layers of this component stack.
First, there are a few consolidated efforts to decompose QM software into more independent components such as the PySCF program \cite{pyscf} implemented in the highly modular Python programming language
 and the ESL library bundle \cite{esl} that encapsulates most of the functionality of a plane-wave pseudopotential QM program.
Next, there are a variety of interfacing layers to enable existing software to function as interoperable components such as the high-level SEAMM environment \cite{seamm}
 that enables the construction of multi-program workflows as graphical flowcharts and the low-level MDI library \cite{mdi} for run-time inter-program communication.
Data is the most ubiquitous interface between programs and users, and there is active movement towards findable, accessible, interoperable, and reusable (FAIR) data standards
 throughout atomistic simulation \cite{fair} and important specific instantiations like the Basis Set Exchange \cite{bse} for Gaussian basis set data
 and the QCSchema \cite{qcarchive} that enables interoperable use of quantum chemistry software in databases and workflow systems.
A growing number of critical components of QM software now have software libraries, including Gaussian AO integrals \cite{libint}, DFT exchange-correlation functionals \cite{libxc},
Hamiltonian matrix solvers \cite{elsi}, molecular geometry optimizers \cite{libopt}, and self-consistent field (SCF) solvers \cite{libscf}.

%P2.6 - SCF as an example reliability outlier
While interfaces and software are major parts of developing atomistic simulation components, refinement of the underlying algorithms, methods, and even formalism may also be important in some cases.
Low-level problems in a component stack can be mitigated by interface and software design such as error handling, but they cannot be fixed without resolving the problem at its lowest level.
Without some active form of error correction, reliable QM software needs reliable components.
By far, the least reliable part of any QM software is the SCF cycle, because development  has focused on algorithms such as direct inversion in the iterative subspace (DIIS) \cite{diis}
 that reduce costs for typical problem instances in chemistry rather than reduce the cost or prevalence of difficult problem instances that suffer from convergence pathologies.
SCF solvers have been attracting increasing attention from applied mathematicians \cite{scf_math},
 who have demonstrated and proven the increased reliability of direct minimization SCF solvers over mixing algorithms like DIIS.
However, SCF solver algorithms based on direct minimization tend to be slower for typical problem instances, and it is not yet clear if this is a temporary or fundamental limitation.
Reliable SCF solvers may also need active electronic orbitals with fractional occupations in some problem instances \cite{scf_fraction},
 which is not directly compatible with the strict partition of electronic orbitals into occupied and virtual in many quantum chemistry programs.
Ideally, developer culture in atomistic simulation should align software and method development concerns in the transition towards more component-centric development,
 with software support activity focused on identifying practical priorities for future method development
  and software maintenance activity focused on implementing new methods and integrating new knowledge as they emerge.

%S2.1
\subsection{SQM model deconstruction}

%P2.1.1 - SQM models as 3 components: model Hamiltonian, correlation model, & approximate solver
While the historical development of SQM models has been monolithic, their many design choices can be tabulated and compared \cite{sqm_roadmap}.
In addition to differing in how they are fit and what they are fit to, these design choices primarily cluster into three components: a model Hamiltonian, a total energy model, and a solver algorithm.
Improved modularity of these components is important to understanding their individual costs and errors.
SQM models are intentionally cost-limited approximations, which makes it particularly important to understand how costs and errors are budgeted between their different components.
A lack of such understanding can imbalance costs and errors, reducing the overall effectiveness of an SQM model.
Historical examples of this include the limited effectiveness of combining second-order many-body perturbation theory with MNDO-family models \cite{mndoc},
 and the persistent failure of MNDO-family models to describe the electrostatics of hydrogen bonds \cite{mndo_failures} without explicit classical corrections \cite{mndo_hbond}.
Building SQM models from more modular components also enables a separation of concerns between improving components and designing new SQM models that use them.

%P2.1.2 - Compare & contrast w/ Hamiltonian downfolding
In isolation, the fitting of a model many-electron Hamiltonian in SQM has the same goal as more formal QM methods for transforming \textit{ab initio} Hamiltonians into simpler model forms \cite{downfold}.
The main difference is that SQM model Hamiltonians are parameterized to fit reference data directly rather than carrying out an explicit many-body operator transformation.
These research activities have a lot to learn from each other, as SQM has been successfully deploying specific parameterized families of model Hamiltonians for decades
 while many-body transformations offer deeper insight into the effectiveness of different Hamiltonian model forms.
More specifically, efforts to increase the accuracy of SQM models are now going beyond a minimal AO basis and following the popular quantum chemistry prescription of
  including split-valence and polarization functions \cite{dzp_sqm}.
However, increasing the number of AOs per atom tends to increase the condition number of the AO overlap matrix,
 which may have more severe consequences for SQM models than \textit{ab initio} QM calculations.
A possible alternative might be to revisit EPM model forms \cite{epm} to construct SQM model Hamiltonians with both a minimal AO basis
 and a set of smooth, uniform basis functions with a similar number of functions per atom as a larger AO basis.
The resulting hybrid SQM model Hamiltonians would have new classes of matrix elements to parameterize such as matrix elements between AOs and low-degree polynomials.

%P2.1.3 - Total-energy models on a budget
The choice of total energy components for SQM models is slowly expanding with increasing progress in QM method development.
The MNDO-family models in MOPAC \cite{MNDO} are based on Hartree-Fock calculations with total energies fit to reproduce experimental heats of formation
 because that was the efficient level of theory and accessible reference data available in its time.
The later DFTB models \cite{dftb} used expansions around an atomic limit to construct total energy models that approximate \textit{ab initio} DFT calculations,
 which then served to supplement experimental reference data.
As independent components, a model Hamiltonian and total energy model should both be as accurate as possible in isolation,
 although it may still be beneficial to reparameterize model Hamiltonians to compensate for errors in the accompanying total energy model.
Fast algorithms can enable more accurate total energy models in cost-constrained SQM models, and recent innovations include
 cubic-scaling algorithms for the random-phase approximation (RPA) and second-order M{\o}ller-Plesset perturbation theory \cite{rpa}.
While SQM models tend to focus on ground-state properties, there are also advances in total energy models for excited states such as the use of quasiparticle methods with tight-binding models \cite{gw}.
Because of the broad interest in the quantum many-body problem, new total energy components may come from research areas outside of atomistic simulation,
 such as quantum information theory, nuclear physics, or high-energy particle physics.

%P2.1.4 - Approximate solvers & their consequences
To achieve a low computational cost, SQM models may need to use solver algorithms that either introduce additional approximations and errors
 or reduce the domain of applicability, or else wait for future methodological advances.
There was a substantial amount of research in the 1990s on linear-scaling QM solvers \cite{linear_scaling},
 and SQM models were the first beneficiaries of this research because they could effectively use these solvers for smaller system sizes than \textit{ab initio} QM applications.
The MOZYME solver in MOPAC \cite{mozyme} was one of the earliest linear-scaling solvers with a practical deployment because its application domain
 is restricted to closed-shell systems with identifiable formal atomic charges and bond orders.
More generally applicable linear-scaling solvers are now available to SQM models \cite{linear_scaling2}, but they introduce errors that must be budgeted into the overall SQM model.
There is a broad range of linear-scaling solvers available to SQM models, with the most efficient and approximate solvers having natural connections to bond-order potentials used in MM methods \cite{bop}.
While linear-scaling solvers with both high accuracy and general applicability do not yet exist, progress is slowly being made in numerical linear algebra research towards that end.
For example, the pole expansion and selected inversion (PEXSI) method \cite{pexsi} is a quadratic-scaling QM solver that utilizes modern advances in sparse matrix factorization and is not restricted to closed-shell systems.
Applied mathematicians have known for over three decades how to compress the free electron propagator matrix with fast multipole methods \cite{hfmm},
 but those fundamental advances have still not been generalized to electrons in inhomogeneous molecules and materials.

%S2.2
\subsection{Statistical model selection}

%P2.2.1 - Summary of model selection
Once its components have been selected, an SQM model is constructed by fitting its parameters to reference data.
Typically, the fitting process is carried out as a nonlinear least-squares optimization \cite{sqm_fit}, such as is implemented in the PARAM fitting program that accompanies MOPAC.
However, this process does not provide any operational meaning to a model's goodness of fit,
 nor does it provide feedback on the selection of model components or protection against overfitting.
All of these benefits can be obtained by applying modern statistics and its formalism for model selection to the construction of SQM models \cite{model_selection}.
This is a maximalist approach that comes with many technical requirements, but it provides a precise statistical meaning to constructing the SQM model that is most likely to succeed at a simulation task,
 provided that a precise statistical description for a desired distribution of tasks is constructed first.

%P2.2.2 - Deeper understanding of use & cost
In the maximalist interpretation, a set of reference data corresponds to independent samples from a target distribution of atomistic simulation tasks.
For example, the experimental heat-of-formation data used to parameterize the PM$x$ models in MOPAC are direct samples from the distribution of molecules that are of experimental interest.
QM method development has been steadily aggregating its own large reference data sets such as MGCDB84 \cite{qm_data} but it is less clear how they relate to task distributions of practical applications.
Reference data serves as more of a satisfiability test in QM method development because such methods are expected to have a high degree of transferability,
 therefore they can be expected to be accurate in general if they are accurate in a sufficient number of arbitrary examples.
The effectiveness of SQM models depends on the similarity of testing and training sets, such as in the case of GFN$x$ models performing better than PM$x$ models for conformer energy prediction \cite{conform}
 because GFN$x$ models are more directly trained on intermolecular interaction energies.
With the ongoing exponential growth in the use of atomistic simulation methods and software,
 it is becoming increasingly important to understand this use better and characterize it with a set of task distributions corresponding to the most prevalent patterns of use.
The work of constructing such task distributions is distinct from QM method development or SQM model fitting, but it can help to target those activities more effectively at real demand.
Even \textit{ab initio} QM method development is dependent on task distributions, with molecular applications dominated by the Gaussian AO formalism
 and materials applications dominated by the plane-wave pseudopotential formalism.

%P2.2.3 - Basis for ranking & commoditization
An additional benefit of having well-developed task distributions and success metrics is that they can be used to score and rank different methods and software.
Rankings can offset the confusion and decision paralysis of a large number of choices, such as choosing Gaussian AO basis sets or DFT exchange-correlation functionals.
Leaderboard websites for ranking atomistic simulation methods and software are already beginning to emerge \cite{leaderboard}.
If rankings include the average computational cost alongside the success metric, then a Pareto front of cost versus accuracy can be constructed for the available software.
Significant gaps in these Pareto fronts can then identify deficiencies in atomistic simulation capabilities and useful targets for new method and software development.
Such a reduction of diverse methods and software to cost-accuracy curves for a set of task distributions is a step towards the commoditization of atomistic simulation.

%S3
\section{Hamiltonian model forms}

%P3.1 - Lack of standards in SQM models
SQM and QM method development have been operating at different activity levels that have resulted in different standard practices.
Each popular SQM model family has been developed independently by a relatively small group of researchers and implemented in its own software with its own standards.
Only very recently has this situation begun to change with software such as Sparrow \cite{sparrow} and ULYSSES \cite{ulysses} 
 providing implementations of multiple SQM model families in a common code base with a common interface.
In contrast, there are many QM software programs that implement either the Gaussian AO or plane-wave pseudopotential formalisms.
Within those formalisms, there are many distinct instances of standard components such as Gaussian AO basis sets, pseudopotentials, and DFT exchange-correlation functionals.
To accommodate and encourage this level of activity, these components have standard formal structures and often standard data formats, which enables their large-scale aggregation
 such as in the Basis Set Exchange website \cite{bse} or the Libxc software library \cite{libxc}.
More standards in SQM model structures could similarly encourage more activity, and enable more interoperability between different SQM development efforts and between QM and SQM efforts.
In particular, a common model form for many-electron Hamiltonians could enable QM software to apply the same methods to both QM and SQM Hamiltonians.
Although many QM software programs contain implementations of SQM models, they are usually isolated components that are incompatible with most of the other QM features.

%P3.2 - Summarize results of section
Practically, a more standard and interoperable form for both QM and SQM Hamiltonians corresponds to a standard form for the important matrix elements that define these Hamiltonians.
Here, I consider the specific case of AO-based matrix elements that are relevant to all popular SQM models and all QM software using the Gaussian AO formalism.
To enable parameterized models with a manageable number of parameters, the structured form of SQM matrix elements must be designed to minimize the amount of data
 whereas QM matrix element structure is usually designed to minimize approximation error.
In particular, the structured form that I propose here is not exactly compatible with standard factorizations of four-center Coulomb integrals in QM,
 but it is a systematically improvable approximation that has computational benefits inspired by the efficiency of SQM calculations.

%S3.1
\subsection{Two-center matrix elements}

%P3.1.1 - Slater-Koster formalism
The standard two-center AO matrix elements in SQM and QM calculations have the general form
\begin{equation}
 K_{p,q} = \int d\mathbf{r} d\mathbf{r'} \phi_p(\mathbf{r}) K(\mathbf{r}, \mathbf{r'}) \phi_q(\mathbf{r}'),
\end{equation}
 where $K$ is a translationally and rotationally invariant operator such as the Coulomb interaction, kinetic energy, or identity, and $\phi$ are real-valued AOs.
It is notationally useful to embed information about the AOs into their labels, $p = (\mathbf{r}, \tau, l, m)$ for a center $\mathbf{r}$, type $\tau$, angular quantum number $l$,
 and magnetic quantum number $m$, and similarly $q = (\mathbf{r}', \tau', l', m')$.
The Slater--Koster tight-binding formalism \cite{slater_koster} established that these matrix elements can be decomposed into a universal form,
\begin{equation} \label{factor_matrix_elements}
  K_{p,q} = \sum_{M = -L}^{L} R^{l}_{m,M}(\mathbf{u}) K_{|M|}^{\tau, \tau'}(|\mathbf{r}-\mathbf{r'}|) R^{l'}_{m',M}(\mathbf{u})
\end{equation}
for $L = \min\{l,l'\}$ and $\mathbf{u} = (\mathbf{r}-\mathbf{r'})/|\mathbf{r}-\mathbf{r'}|$.
$R_{m,m'}^l(\mathbf{u})$ are matrix elements between real spherical harmonics $Y_{l,m}$ of standard alignment to $\mathbf{z}=(0,0,1)$ and $Y_{l,m'}$ aligned to $\mathbf{u}$,
 which can be evaluated using fast, stable recursive formulas \cite{rotate}.
$K^{\tau, \tau'}_m(r)$ are real-valued Slater--Koster model parameters with a symmetry, $K^{\tau, \tau'}_m(r) = (-1)^{l+l'}K^{\tau', \tau}_m(r)$, that reduces the set of independent parameters to $0 \le m \le l \le l'$.

%P3.1.2 - Benefits of flexibility
The main benefit of using the Slater--Koster formalism is its flexibility.
$K^{\tau, \tau'}_m(r)$ can be evaluated as integrals of AOs in a standard pose with $r\mathbf{z}$ as the distance vector from $\phi_q$ to $\phi_p$,
 or by interpolation from values in a Slater--Koster table, or from a parameterized SQM formula.
Typically, they are truncated beyond a cutoff radius, either to zero for local matrix elements or to a multipole-multipole interaction for Coulomb matrix elements.
While this formalism can be used with orthogonal AOs, the use of non-orthogonal AOs with overlap matrix elements $S_{p,q}$ is empirically known to improve transferability \cite{transferability,transferability2},
 and it is formally required to maintain $l$ and $m$ as good quantum numbers.
This flexibility also makes it easier to reconsider and transition from old SQM model forms such as the Klopman--Ohno electrostatic interaction,
 which corresponds to charge distributions with unphysical algebraic tails that significantly complicate periodic electrostatics \cite{ohno}.
It also accommodates a broad range of parametric complexity, from general distance dependence for each pair of elements and shells
 to a single Slater orbital exponent or Klopman--Ohno radius for each element and shell as in the MNDO-family models \cite{MNDO}.
Having a single length-scale parameter for each element and shell is appropriate for models that are fit mostly to equilibrium structures with a limited variety of nearest-neighbor distances.

%S3.2
\subsection{Four-center matrix elements}

%P3.2.1 - Auxiliary basis sets
The standard four-center Coulomb AO matrix elements in SQM and QM calculations have the general form
\begin{equation}
 V_{pq,rs} = \int d\mathbf{r} d\mathbf{r'} \phi_p(\mathbf{r}) \phi_q(\mathbf{r}) \frac{1}{|\mathbf{r} -\mathbf{r'}|} \phi_r(\mathbf{r}') \phi_s(\mathbf{r}'),
\end{equation}
 and density-fitting methods use two-center Coulomb matrix elements in an auxiliary AO basis,
\begin{equation}
 V_{\mu,\nu} = \int d\mathbf{r} d\mathbf{r'} \rho_\mu(\mathbf{r}) \frac{1}{|\mathbf{r} -\mathbf{r'}|} \rho_\nu(\mathbf{r}'),
\end{equation}
 to approximate them using the factored form
\begin{equation} \label{factor}
 V_{pq,rs} \approx \sum_{\mu,\nu} X_{p,q}^\mu V_{\mu,\nu} X_{r,s}^\nu.
\end{equation}
It is popular in quantum chemistry to fit a general three-center $X_{p,q}^\mu$ to minimize errors in a chosen auxiliary AO basis.
In SQM models, a general $X_{p,q}^\mu$ has an intractable number of parameters to fit, and it is more practical to restrict $\mu$ to the same atoms as $p$ and $q$.
Such two-center density fitting has a long history \cite{old_local_ri} and modern versions that attain high accuracy \cite{local_ri}.

%P3.2.2 - One-center charge matching
In the one-center case with $p$ and $q$ located on the same atom, $X_{p,q}^\mu$ is also restricted to non-zero values only for $\mu$ on that atom.
For SQM models, these one-center $X_{p,q}^\mu$ tensors are arbitrary parameters within the sparsity pattern of non-zero Clebsch--Gordan coefficients.
The only important constraint is that the auxiliary AO approximation should conserve electronic charge, which corresponds to the condition
\begin{equation} \label{charge_condition}
 \sum_{\mu} X_{p,q}^\mu Q_\mu = S_{p,q},
\end{equation}
 where $S_{p,q}$ is the one-center AO overlap matrix, which is typically the identity matrix, and $Q_\mu$ are the total charges of the auxiliary AO basis functions.
For explicit function approximations, this fit can be achieved by a local approximation using the Coulomb metric with multipole moment constraints \cite{multipole_match}.

%P3.2.3 - Tertiary basis sets
In the two-center case with $p$ and $q$ on different atoms, local density fitting with the Coulomb metric requires a second approximation to limit the number of SQM parameters,
\begin{equation}
 \int d\mathbf{r} V_\mu(\mathbf{r}) \phi_p(\mathbf{r}) \phi_q(\mathbf{r}) \approx \sum_{r} Z^\mu_{p,r} S_{r,q},
\end{equation}
 where $V_\mu$ is the electrostatic potential of $\rho_\mu$, $S_{p,q}$ is the two-center AO overlap matrix, and $Z^\mu_{p,q}$ is another one-center tensor of parameters.
This separates a two-center, three-index integral into a sum over products of two-center, two-index terms and one-center, three-index terms.
Interpreted as a function approximation, products of $V_\mu$ and $\phi_p$ on the same atom are approximated in the primary AO basis of that atom.
For explicit function approximations, a larger tertiary AO basis could be introduced here to reduce fitting errors.
The two-center form of $X_{p,q}^\mu$ with conserved charge is then
\begin{subequations} \label{2center}
\begin{align}
 X^\mu_{p,q} &= \sum_{\nu} \Lambda_{\mu,\nu} (Q_\nu w_{p,q} + W^\nu_{p,q}), \\
 W^\mu_{p,q} &= \sum_{r} (Z^\mu_{p,r} S_{r,q} + Z^\mu_{q,r} S_{r,p} ), \\
 w_{p,q} &=  \frac{S_{p,q} - \sum_{\mu,\nu} Q_\mu \Lambda_{\mu,\nu} W^\nu_{p,q}}{\sum_{\mu,\nu} Q_\mu \Lambda_{\mu,\nu} Q_\nu},
\end{align}
\end{subequations}
 where $\Lambda_{\mu,\nu}$ is the matrix inverse of the corresponding two-center matrix block of $V_{\mu,\nu}$.
Explicit function approximations can constrain higher multipole moments to reduce errors \cite{multipole_match}.

%P3.2.4 - Benefits of the more factored structure
Overall, this SQM model of four-center Coulomb integrals depends only on two-center, two-index parameters $V_{\mu,\nu}$ and $S_{p,q}$
 and one-center, three-index parameters $X^{\mu}_{p,q}$ and $Z^{\mu}_{p,q}$.
This factored tensor structure maintains a modest parametric complexity and enables the use of fast solver algorithms based on intermediate tensor contractions \cite{rpa}.
It reduces to the NDDO approximation \cite{nddo} in the case of orthogonal orbitals,
 and the NDDO approximation is the main reason why SQM calculations are more efficient than QM calculations in a minimal AO basis.
The NDDO approximation uses orthogonal orbitals to eliminate interatomic monopole moments, but it is fundamentally incapable of accounting for higher-order interatomic multipole moments.
While the NDDO approximation is reasonably accurate for individual Coulomb integrals,
 large overall errors can accumulate from a large number of individual integral errors, and increasing the number of AOs does not reduce these errors \cite{nddo_error}.
This more general form retains the computational benefits of NDDO while enabling a systematic reduction of errors by expanding the number of primary and auxiliary AOs.
Besides the NDDO approximation in the MNDO-family SQM models, most popular SQM models retain only a monopole approximation of the Coulomb interaction
 except for the GFN2-xTB model \cite{gfn2}, which uses a multipole approximation up to quadrupoles with parameterized interatomic multipole moments and structure similar to Eq.\@ (\ref{2center}).

%S4
\section{Ongoing work and challenges}

%P4.1 - Note the diversity of ongoing work (both a good and bad thing)
While method development in SQM is presently less active than QM and MM, there is still some notable ongoing work pursuing a variety of technical directions.
In particular, there is a concentration of research activity in combining machine learning (ML) models with SQM models.

%P4.2 - Three-body TB
The NRL tight-binding model \cite{nrltb} was the most substantial open effort to build SQM models for materials, but it was limited to unary systems for 53 elements of the periodic table.
The ThreeBodyTB.jl software \cite{three_body} is ongoing work to extend the SQM modeling of materials to binary systems comprised of 65 elements.
Its expanded and refined SQM model contains three-body Hamiltonian corrections and self-consistent atomic charges to improve transferability,
 and it is fit to total energies and band structures of DFT reference data.

%P4.3 - NOTCH
There is ongoing work to incorporate QM reference data more directly into SQM models through a more explicit Hamiltonian downfolding process
 into a Natural Orbital Tied Constructed Hamiltonian (NOTCH) form \cite{notch}.
Similar efforts date back to Dewar's post-AM1 work on the SAM1 model \cite{sam1},
 and NOTCH is an increase in technical sophistication that replaces more of the empirical fitting with a formal transformation between basis sets.
The NOTCH form includes a complete set of two-center Coulomb integrals instead of using the NDDO approximation, but it still neglects three-center and four-center Coulomb integrals.
So far, NOTCH has only been tested on diatomic molecules, which are not sensitive to these neglected Coulomb integrals,
 and more transferable versions of NOTCH may need to approximate these integrals rather than neglect them.

%P4.4 - DZP
Other ongoing work in SQM model construction is drawing from a larger set of components such as expanded AO basis sets with split-valence and polarization functions
 and approximate, fixed-cost SCF solvers \cite{dzp_sqm}.
Such work highlights the need to better understand sources of error in SQM models to budget costs and errors more effectively with an increasing number of model design choices.
The larger basis set increases cost and accuracy while the fixed-cost SCF solver prevents systems with slow SCF convergence from becoming cost outliers,
 but it may convert these cost outliers into error outliers and further broaden the tails of error distributions.
Notably, this more modern SQM model form now has enough control over sources of error to benefit from larger basis sets, unlike older efforts to construct SQM models beyond a minimal basis \cite{thiel_dz,other_dz}.

%P4.5 - XTB software ecosystem: SQMBox, tblite, & dxtb
There is also ongoing work to refine SQM software in addition to SQM models, particularly within the GFN family of models.
The tblite library \cite{tblite} is a result of work to increase the modularity of xTB, which provides convenient access to GFN$x$ model Hamiltonians for other SQM and QM software.
The dxtb framework \cite{dxtb} is a Python implementation of GFN$x$ models using automatic differentiation to support refitting of model parameters to custom reference data.
The SQMBox library \cite{sqmbox} tightly integrates GFN$x$ models into the TeraChem software through the elementary operation of mapping from a density matrix to a Fock matrix,
 which may also enable integration with other QM software of a similar design.

%S4.1
\subsection{Role of machine learning}

%P4.1.1 - typical uses of ML in SQM
Because ML research is very popular now and in close technical proximity, most of the ongoing work in SQM research is based on ML methods or software.
This work is mostly clustered into two categories of activity.
The first category of work uses ML models as an interatomic potential to correct the potential energy surface produced by an existing SQM model, thus forming a more accurate hybrid model.
The second category of work integrates ML machinery directly into SQM models as a systematic framework for adding and fitting new model parameters, usually as part of the SQM model Hamiltonian.
In either case, ML components are increasing the flexibility of SQM models, which enables them to fit more reference data and also improve the accuracy of such fits.

%P4.1.2 - fitting post-SQM classical ML corrections
In the first category, the development of established SQM model families is continuing through the addition of new ML corrections to the total energy, often replacing simpler corrections based on physical models.
The AIQM1 model \cite{aiqm1} continues the development of orthogonalization-corrected MNDO-family models by replacing dispersion corrections with a modified version of the ANI-1x model \cite{ani}.
The PM6-ML model \cite{ml_pm6} continues the development of the PM$x$ model family by replacing hydrogen-bond corrections with a TorchMD-NET potential \cite{torchmd}.
The QD$\pi$ model \cite{qdpi} corrects a DFTB model using a DeepPot-SE potential \cite{deepmd}.
In some cases, the combination of SQM and ML models is inspiring new ML model forms such as the ChIMES cluster expansion \cite{chimes},
 which has been used as a flexible alternate form for the repulsive correction in DFTB models.

%P4.1.3 - using ML machinery to fit the quantum parts of SQM models
In the second category, there is more diverse research activity because there are many ways to combine ML and SQM models.
Static parameters of an SQM model can be replaced by environment-dependent parameters generated by an ML model,
 as demonstrated by a combination of the PM3 model and a deep neural network \cite{deep_sqm},
 which is a further increase in model flexibility relative to older charge-dependent SQM parameters \cite{charge_sqm}.
The sophisticated optimization machinery used to fit ML models can be adapted to SQM model forms,
 such as with the direct optimization of Slater-Koster parameters in a spline representation \cite{ml_sqm}.
ML software frameworks can be used to implement existing SQM models and prepare them for combined optimization alongside ML corrections,
 which is the approach of the TBMaLT software \cite{tbmalt}.
While most of this activity is focused on AO-based models, there is even some renewed interest in EPM models,
 with the use of equivariant neural networks to generate more transferable EPM model parameters \cite{ml_epm}.

%P4.1.4 - caveats of ML usage
The research activity in combining ML and SQM models is likely to increase for the foreseeable future, but it should remain mindful of two important caveats.
First, the two SQM model families that have achieved the widest coverage of the periodic table, PM$x$ \cite{pm7} and GFN$x$ \cite{xtb},
 use matrix elements with monoatomic parameters and simple distance dependence in their Hamiltonian model forms.
The general form of Slater--Koster matrix elements allows for diatomic parameters and arbitrary distance dependence,
 but only recently has there been enough QM reference data to fit these more general forms \cite{three_body}.
Monoatomic and diatomic parameters correspond to linear and quadratic growth in the number of model parameters with the number of elements covered in the periodic table.
In general, ML models are likely to have a combinatorial growth of parameters if they maintain a constant accuracy while increasing elemental coverage,
 and it will be challenging to generate enough reference data to fit this growing number of parameters.
Finally, ML models are often considered to have a negligible computational cost relative to SQM models,
 but these costs are not always well separated and might eventually cross as ML models become more complicated.
For example, PM6-ML calculations of large proteins using the fast MOZYME solver spend 13\% of their total computational cost on evaluating their ML correction \cite{ml_pm6}.
This is part of a general trend in MM, where the cost of interatomic potentials has been steadily growing as their complexity and accuracy have both increased \cite{mm_cost}.
If the cost of the ML component exceeds the cost of the SQM component in a hybrid model,
 then a better cost-accuracy balance might be achieved by switching to a more accurate and expensive SQM component.

%S4.2
\subsection{Hydrogen cluster example}

%Figure 3 - H model example
\begin{figure*}[!t]
\includegraphics{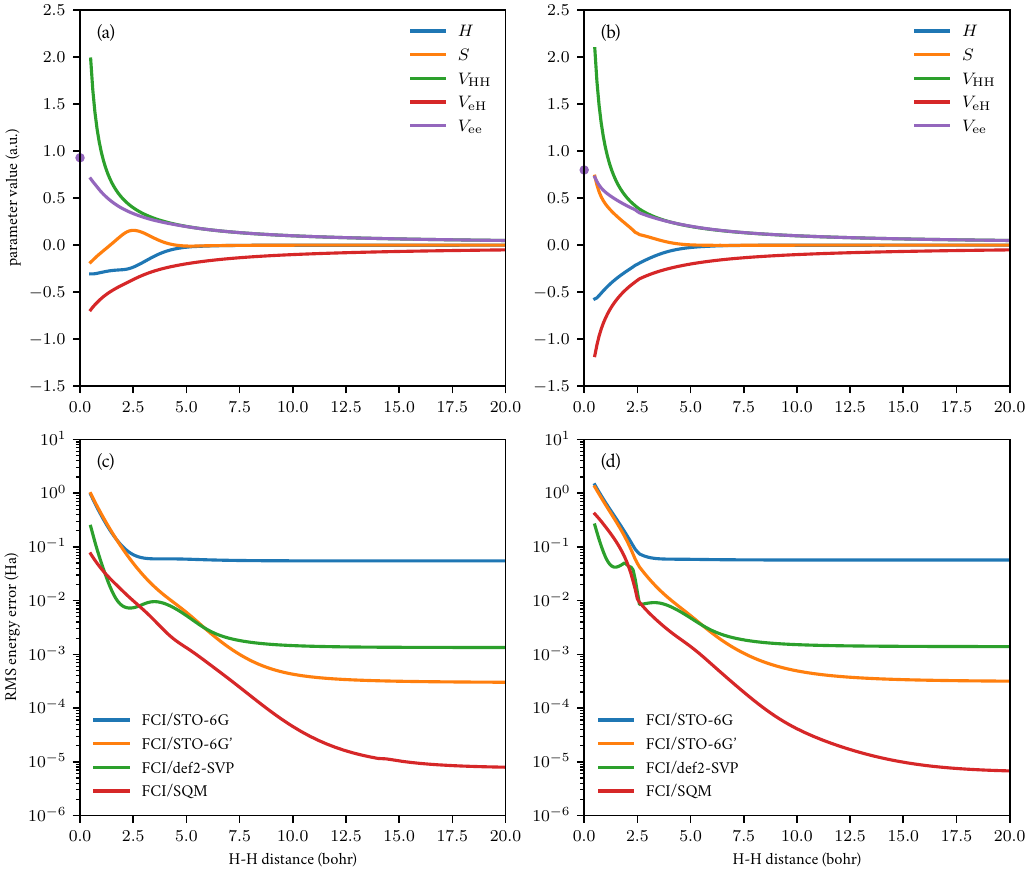}
     \caption{Optimized Slater--Koster model parameters (a,b) and RMS energy errors relative to FCI/def2-QZVPP reference data (c,d) for the two tasks of
     calculating the ground-state energy of each distinct charge and spin sector (a,c) and calculating all distinct stationary-state energies in the single-occupancy sector (b,d) of H$_2$ and symmetric H$_3$.
}\label{fit_fig}
\end{figure*}

%P4.2.1 - Benefits of error analysis for components
Before concluding this paper, I consider a simple numerical example that shows the value of understanding errors in individual SQM model components.
Hydrogen is the most prevalent element in biomolecular simulations, so the cost and accuracy of SQM models in biomolecular applications is sensitive to how hydrogen is modeled.
While most AO-based SQM models consider only one $s$ orbital on each hydrogen atom, there are models that add a second $s$ orbital \cite{xtb}, or a shell of $p$ orbitals \cite{pmo}, or both \cite{dzp_sqm}.
However, it is difficult to separate the effect of that specific design decision from all of the other differences in model design and fitting.
Here, I examine some of the limits of an SQM model of hydrogen with a minimal AO basis.

%P4.2.2 - Scope of numerical experiment
I make several simplifying assumptions in this example.
First, I consider a model Hamiltonian using Slater--Koster parameters and factored four-center Coulomb integrals from Eq.\@ (\ref{factor})
 with a single $s$-type primary and auxiliary AO on each hydrogen atom.
This minimal basis simplifies the density fitting structure, reducing the model to two one-center parameters---orbital energy and electronic Coulomb self-energy---and five two-center parameters---overlap ($S$), one-body Hamiltonian ($H$), and electron-electron ($V_{\mathrm{ee}}$), electron-proton ($V_{\mathrm{eH}}$),
  and proton-proton ($V_{\mathrm{HH}}$) Coulomb interactions.
I use the exact AO energy for an isolated atom, leaving one one-center parameter and five two-center parameters.
Second, I model electron correlation exactly using full configuration interaction to isolate the Hamiltonian model errors.
Third, I restrict the reference data to the hydrogen dimer and the symmetric hydrogen trimer so that the Slater--Koster parameters can be fit independently for each interatomic distance.
I then generate FCI/def2-QZVPP reference data using PySCF \cite{pyscf} for all stationary electronic states in the single-occupancy sector of the energy spectrum
 corresponding to the number of electronic configurations in a minimal AO basis with up to one electron per atom.

%P4.2.3 - Overview of results
I consider two versions of this SQM model that are least-squares fits to two different simulation tasks.
The first task is to evaluate the 10 ground-state energies for each distinct spin and charge sector, and
 the second task is to evaluate all 23 distinct stationary-state energies in the single-occupancy sector.
In both cases, I fit the Coulomb self-energy to minimize the root-mean-square (RMS) energy error at 1.4 bohr, near the equilibrium bond length of H$_2$.
The errors and optimized model parameters for these two tasks are shown in Fig.\@ \ref{fit_fig}, and the errors are compared against \textit{ab initio} QM calculations with small basis sets.
The minimal-basis SQM model is uniformly better than the minimal-basis QM results, even after the standard STO-6G basis for hydrogen is rescaled to match the 1s orbital of the isolated atom (STO-6G').
QM results in a polarized, split-valence AO basis (def2-SVP) are better than the SQM model in the bonding region near 1.4 bohr,
 showing that semiempirical fitting of a minimal basis is not always enough to match the variational flexibility of a larger basis.
The SQM errors are more sensitive than the QM errors to the target task distribution, as a limited number of parameters are more frustrated by fitting to a broader set of tasks.
Even for the narrower set of tasks, a minimal-basis SQM model is insufficient to achieve the typical chemical accuracy target of 0.0016 Ha for small interatomic separations,
 and a larger basis is warranted for chemically accurate SQM models.

%P4.2.4 - Path to a systematic construction of a core SQM model
The fitting of SQM models can only practically find local minima in parameter space, so a good initial guess is needed to have a good chance of finding the global minimum.
Fitting an accurate model for hydrogen may be a good starting point for fitting parameters of other elements based on accurate reference data for small, hydrogen-decorated atomic clusters
 containing pairs of heavier elements at fixed interatomic distances.
With such data available, initial guesses for Slater--Koster parameters could be fit independently for each pair of elements at each interatomic separation
 after independently fitting the parameters governing the interaction between each element and hydrogen.
The separation of SQM model fitting into a large number of small, independent optimization problems could reduce technical barriers in future SQM research.
Similarly, the widespread use of pseudopotentials in \textit{ab initio} QM was enabled by designing them around simple, independent fitting processes for isolated atoms \cite{modern_pseudopotentials}.

%S5
\section{Conclusions}

%P5.1 - Inherent significance of SQM
This paper has emphasized the pragmatic nature of SQM method and software development, which is a faithful representation of its historical emphasis.
However, SQM method development might also be viewed idealistically as an operational exploration of the boundary between classical and quantum descriptions of physical reality
 that seeks to strip away all but the most essential QM components in descriptive models of atomistic systems, to replace computation with data refined from prior knowledge.
SQM models operate at a transferability limit of atomistic modeling, beyond which simpler MM models must either substantially narrow their domain of applicability
 or substantially increase their parametric complexity to maintain accuracy.
Perhaps in the future, these idealistic considerations might complement its historical pragmatism to increase the overall level of academic interest in SQM research.

%P5.2 - Acknowledgement of futility
As evident in the collapsing market share of SQM software over the last three decades, development and application of atomistic simulation has substantially polarized
 towards the extremes of high-accuracy QM and low-cost MM.
The rapidly growing popularity of ML and quantum computing research might be interpreted as a further polarization, with QM methods now using expensive quantum computing hardware
 in the pursuit of higher accuracy \cite{quantum_comp} and MM methods now replacing even more physical modeling with data-driven ML models.
Atomistic method development would perhaps be better served by an interconnected pipeline from high-cost, high-accuracy QM methods
 to SQM models of lower accuracy or narrower applicability before fragmenting into a constellation of special-purpose MM models.
There have already been multiple efforts over the last three decades to encourage and advocate
 for more SQM research \cite{motive1,motive2,motive3,motive4,motive5,sqm_roadmap} that have not arrested this collapse.
I have argued in this paper for the enduring relevance of SQM to atomistic simulation, regardless of its decline.

%P5.3 - SQM should be competitive in a future cost-sensitive market
Even without a substantial increase in academic activity, the steadily growing market for atomistic simulation might eventually be large enough
 to support SQM method and software development solely through commercial demand.
While SQM software presently has a very small market share, latent demand is difficult to quantify,
 and new SQM software that is more responsive to demand might eventually increase its market share once again.
Historically, SQM models have mostly been products of academic research that were freely distributed,
 but some SQM models such as the QUASINANO parameter sets \cite{quasinano,quasinano2} are now also being developed for commercial distribution.
A financially sustainable future for SQM method and software development may necessitate a careful partitioning between open and closed components.

\section{Biography}

\begin{center}
\includegraphics[width=5cm]{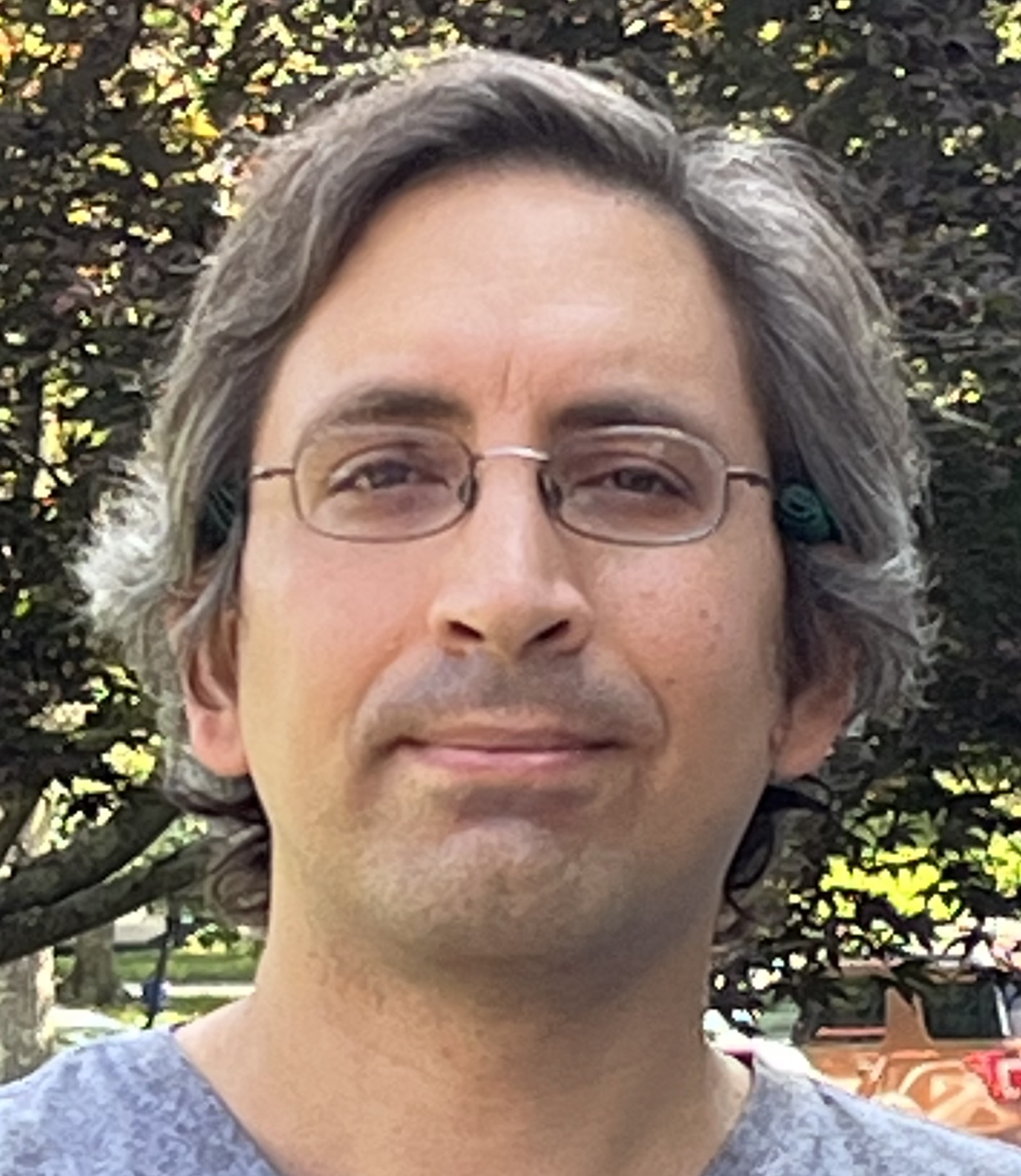}
\end{center}

\textbf{Jonathan E. Moussa} received his doctoral degree in Physics from the University of California, Berkeley in 2008 under the supervision of Prof.\@ Marvin Cohen.
His doctoral work focused on applying density functional theory to predict new phonon-mediated superconductors.
After two postdoctoral positions in computational materials science, at the University of Texas, Austin with Prof.\@ James Chelikowsky and at Sandia National Laboratories with Dr.\@ Peter Schultz,
 he switched fields to quantum information science and converted to a staff scientist at the Labs in 2014.
In 2018, he joined the Molecular Sciences Software Institute as a software scientist and returned to computational materials science and chemistry.
In this role, he led the transition of MOPAC from commercial to open-source software and its ongoing development, maintenance, and support.
His main research interest is the balance between cost and accuracy in algorithms and software for atomistic simulation,
 particularly at the interface between quantum and classical in methods and computers.

\begin{acknowledgement}
J.E.M.\@ thanks Christoph Bannwarth, Paul Saxe, and Jimmy Stewart for useful discussions.
The Molecular Sciences Software Institute is supported by grant CHE-2136142 from the National Science Foundation.
The computational resources used in this work were provided by Advanced Research Computing at Virginia Tech.
\end{acknowledgement}

\begin{suppinfo}

The following files are available.
\begin{itemize}
  \item \texttt{methods.pdf}: Methodology for market analysis and data collection.
  \item \texttt{software.zip}: Software used to generate data and figures.
  \item \texttt{data.zip}: Data used in this paper.
\end{itemize}

\end{suppinfo}


\end{document}